\documentclass[%
 reprint, 
 amsmath,amssymb,
 aps, prl
]{revtex4-2}

\usepackage{graphicx}
\usepackage{dcolumn}
\usepackage{bm}
\usepackage{braket}

\begin{document}

\preprint{APS/123-QED}

\title{\textbf{Signatures of Narrowband and Ultrashort-Pulse-Like Probing in Coherent Anti-Stokes Raman Scattering Using Twin Beams and Sum-Frequency Generation Readout}}%

\author{Yujiro Eto}
\email{eto.yujiro.8r@kyoto-u.ac.jp}
\affiliation{
Frontier Quantum Measurement Unit,
Center for Science Adventure and Collaborative Research Advancement (SACRA),
Kyoto University, Kyoto 606-8502, Japan
}

\author{Hideaki Kano}
\affiliation{
Department of Biosciences and Informatics,
Faculty of Science and Technology,
Keio University,
3-14-1 Hiyoshi, Kohoku-ku,
Yokohama, Kanagawa 223-8522, Japan
}

\author{Masahiro Takeoka}
\affiliation{
Department of Electronics and Electrical Engineering,
Keio University,
3-14-1 Hiyoshi, Kohoku-ku,
Yokohama 223-8522, Japan
}

\date{\today}

\begin{abstract}
We theoretically investigate coherent anti-Stokes Raman scattering (CARS) driven by bright twin beams with sum-frequency generation (SFG) readout. 
Expressing the SFG output in terms of quantum-field correlation functions yields multiple correlation structures with distinct properties associated with different interaction pathways on the bra and ket sides.
Among them, frequency correlations between the CARS probe and SFG gating fields enable the two fields to act collectively as an effective narrowband probe, avoiding probe-induced spectral convolution. 
When this behavior appears on either the bra or ket side, the other retains Raman relaxation information as if probed by an ultrashort pulse.
\end{abstract}

\maketitle


Achieving high temporal and spectral resolution simultaneously remains a central challenge in ultrafast Raman spectroscopy. 
Time–frequency correlations of entangled photon pairs generated by parametric down-conversion (PDC) have motivated several proposals to overcome this trade-off in Raman spectroscopy \cite{10.1021/jz501124a, zhang2022entangled, 10.1021/acs.jpclett.3c02852, fan2024entangled}. 
Related theoretical studies have also explored the use of entangled photons in coherent multidimensional spectroscopy \cite{Ishizaki20, doi:10.1126/sciadv.abj4566, fujihashi23, fujihashi24}. 
However, direct Raman excitation by entangled photon pairs is limited by their low optical intensity. 
Combining entangled photons with classical ultrashort pulses can substantially enhance the Raman excitation rate but requires precise synchronization between the fields. 
A straightforward approach is to excite PDC with ultrashort pulses for synchronization, but the broad bandwidth of the excitation pulse weakens the signal–idler frequency anticorrelation required for high spectral selectivity.

Bright twin beams from high-gain PDC offer both high optical intensity and frequency and temporal correlations \cite{Chekhova:18, Eto:24}, potentially enabling direct excitation of nonlinear optical processes while retaining correlations useful for spectroscopy. 
Indeed, coherent anti-Stokes Raman scattering (CARS) driven solely by bright twin beams has been experimentally demonstrated \cite{7pvj-3dqt}. 
Subsequently, sum-frequency-generation (SFG) readout, which plays a role analogous to single-photon coincidence detection in the low-gain PDC regime, has been experimentally shown to eliminate spectral convolution arising from the probe bandwidth, enabling high spectral resolution.\cite{eto2026allopticaldeconvolutionquantumcrosscorrelation}. 
However, it remains unclear whether contributions that reflect the Raman line shape without probe-induced convolution can also retain information about the temporal evolution of Raman relaxation.

\begin{figure}
\centering
\includegraphics[width=0.45\textwidth]{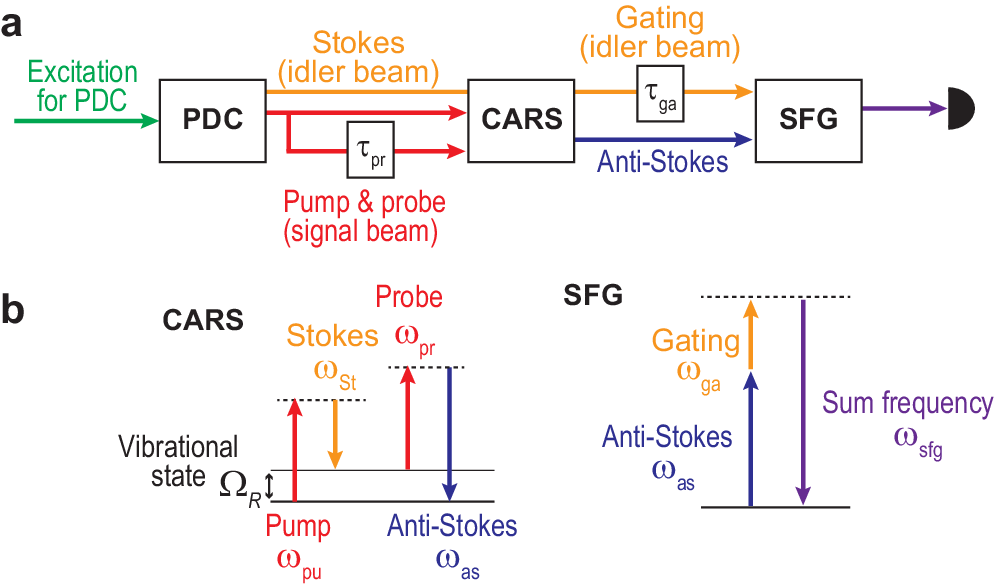}
\caption{
(a) Roles of the signal and idler fields generated by parametric down-conversion (PDC) in coherent anti-Stokes Raman scattering (CARS) and sum-frequency generation (SFG).
(b) Energy-level diagrams for CARS and SFG.
}
\label{fig1}
\end{figure}

In this Letter, we address this question by theoretically investigating the SFG readout of CARS driven by bright signal and idler twin beams, as illustrated in Fig.~\ref{fig1}, 
and show that signatures of both narrowband- and ultrashort-pulse-like probing co-exist through distinct interaction pathways.
We formulate the SFG output as a fourth-order quantum-field correlation function expressed as the inner product of states generated by the CARS–SFG field operators acting along the bra and ket pathways. 
This formulation reveals that different combinations of mode correlations coexist, yielding different types of spectral and temporal Raman information.
Frequency correlations between the CARS probe and SFG gating fields yield a Raman line shape free from probe-induced convolution. 
In this case, the two fields act collectively as an effective narrowband probe, despite their individual spectral bandwidths. 
When this narrowband-like probing behavior appears on either the bra or the ket side, the other side retains Raman relaxation information as if probed by an ultrashort pulse. 
Consequently, distinct interaction pathways give rise to signatures of both narrowband- and ultrashort-pulse-like probing.

We describe the theoretical model. 
The correlated twin beams are generated via PDC. 
In the frequency domain, the PDC field operator is related to the input vacuum field operator through the Bogoliubov transformation
$\hat{a}(\omega) = \mu(\omega)\hat{a}_{v}(\omega) +i\nu(\omega)\hat{a}_{v}^{\dagger}(\Omega_e-\omega)$,
where $\hat{a}_{v}(\omega)$ denotes the vacuum field operator, $\Omega_e$ is the excitation frequency for the PDC process, and $\mu(\omega)$ and $\nu(\omega)$ are the Bogoliubov coefficients characterizing the parametric interaction. 
We consider type-0 PDC and assume the symmetry relations
$\mu(\omega)=\mu(\Omega_e-\omega)$ and $\nu(\omega)=\nu(\Omega_e-\omega)$ \cite{PhysRevA.87.013844}.
The material response is modeled by a single Lorentzian Raman resonance,
$\chi(\omega) = \frac{1}{(\Omega_{R}-\omega)-i\Gamma}$,
where $\Omega_{R}$ is the Raman frequency and $\Gamma$ is the linewidth.

As shown in Fig. \ref{fig1}(a) the signal beam provides both the pump and probe fields, while the idler beam serves as the Stokes field for the CARS process. 
The generated anti-Stokes field is not detected directly. 
Instead, it undergoes SFG with the residual idler beam, which serves as a gating field. 
In this study, we assume that the time delays of the probe and gating fields, denoted by $\tau_{\mathrm{pr}}$ and $\tau_{\mathrm{ga}}$, respectively, can be independently controlled.


The field operator of the generated anti-Stokes light is written as
\begin{equation}
\begin{aligned}
\hat{a}_{\rm as}&(\omega_{\rm as}, \tau_{\rm pr}) \propto
\int^{\Omega_{e}}_{\frac{\Omega_{e}}{2}} d\omega_{\rm pr} 
\int^{\frac{\Omega_{e}}{2}}_{0} d\omega_{\rm St}
\int^{\Omega_{e}}_{\frac{\Omega_{e}}{2}} d\omega_{\rm pu}
\\
&\times
\chi(\omega_{\rm pu}-\omega_{\rm St})
\hat{a}(\omega_{\rm pr})
e^{i\omega_{\rm pr}\tau_{\rm pr}}
\hat{a}^{\dagger}(\omega_{\rm St})
\hat{a}(\omega_{\rm pu})
\\
&\times
\delta(\omega_{\rm pu}-\omega_{\rm St}
+\omega_{\rm pr}-\omega_{\rm as}).
\end{aligned}
\label{eq:CARS_as}
\end{equation}
Here, the subscripts $\rm pu$,  $\rm St$, $\rm pr$, and $\rm as$ denote the pump, Stokes, probe, and anti-Stokes fields, respectively
(see Fig.  \ref{fig1}(b) for the corresponding energy-level diagram).
The factor $e^{i\omega_{\rm pr}\tau_{\rm pr}}$ represents the frequency-dependent phase acquired by the probe field due to the controllable time delay $\tau_{\rm pr}$.
In addition, dispersive phases accumulated by the fields upon propagation through optical elements are neglected, and the corresponding phase factors are taken to be unity.
The field operator of the SFG light generated from the anti-Stokes and gating fields is then written as
\begin{equation}
\begin{aligned}
\hat{a}_{\rm sfg}&(\omega_{\rm }, \tau_{\rm pr}, \tau_{\rm ga})
\propto
\int^{\frac{\Omega_{e}}{2}}_{0} d\omega_{\rm ga}
\int d\omega_{\rm as}\,
\\
&\times
\hat{a}(\omega_{\rm ga})
e^{i\omega_{\rm ga}\tau_{\rm ga}}
\hat{a}_{\rm as}(\omega_{\rm as},\tau_{\rm pr})
\delta(\omega_{\rm ga}+\omega_{\rm as}-\omega_{\rm sfg}).
\end{aligned}
\label{eq:SFG}
\end{equation}
Here, the subscript $\rm ga$ denotes the gating fields.

The SFG spectral intensity, which is proportional to the expectation value of the photon number operator $\braket{\hat{a}^{\dagger}_{\rm sfg}\hat{a}_{\rm sfg}}$, is expressed as
\begin{equation}
\begin{aligned}
S&(\omega_{\rm sfg},\tau_{\rm pr},\tau_{\rm ga}) = \braket{\hat{a}^{\dagger}_{\rm sfg}\hat{a}_{\rm sfg}}
\\
&=
\int_{\frac{\Omega_e}{2}}^{\Omega_e} d\omega'_{\rm pu}
\int_{0}^{\frac{\Omega_e}{2}} d\omega'_{\rm St}
\int_{\frac{\Omega_e}{2}}^{\Omega_e} d\omega'_{\rm pr}
\int_{0}^{\frac{\Omega_e}{2}} d\omega'_{\rm ga}
\\
&\times
\int_{0}^{\frac{\Omega_e}{2}} d\omega_{\rm ga}
\int_{\frac{\Omega_e}{2}}^{\Omega_e} d\omega_{\rm pr}
\int_{0}^{\frac{\Omega_e}{2}} d\omega_{\rm St}
\int_{\frac{\Omega_e}{2}}^{\Omega_e} d\omega_{\rm pu}
\\
&\times
\chi^{*}(\omega'_{\rm pu}-\omega'_{\rm St})
\chi(\omega_{\rm pu}-\omega_{\rm St})
\\
&\times
e^{-i(\omega'_{\rm ga}-\omega_{\rm ga})\tau_{\rm ga}}
e^{-i(\omega'_{\rm pr}-\omega_{\rm pr})\tau_{\rm pr}}
\\
&\times
\bra{0}
\hat{a}^{\dagger}(\omega'_{\rm pu})
\hat{a}(\omega'_{\rm St})
\hat{a}^{\dagger}(\omega'_{\rm pr})
\hat{a}^{\dagger}(\omega'_{\rm ga})
\\
&\qquad
\hat{a}(\omega_{\rm ga})
\hat{a}(\omega_{\rm pr})
\hat{a}^{\dagger}(\omega_{\rm St})
\hat{a}(\omega_{\rm pu})
\ket{0}
\\
&\times
\delta(
\omega'_{\rm pu}-\omega'_{\rm St}
+\omega'_{\rm pr}+\omega'_{\rm ga}
-\omega_{\rm sfg}
)
\\
&\times
\delta(
\omega_{\rm pu}-\omega_{\rm St}
+\omega_{\rm pr}+\omega_{\rm ga}
-\omega_{\rm sfg}
).
\end{aligned}
\label{eq:SFG_spectrum}
\end{equation}
Here, the primed frequency variables are associated with the operators on the ket side, whereas the unprimed frequency variables are associated with those on the bra side.
For classical ultrashort pulses, the corresponding expression is obtained by replacing the field operators with their corresponding complex spectral amplitudes.
The bracket appearing in the integrand of Eq.~(\ref{eq:SFG_spectrum}) represents the inner product between the states associated with the operator sequences on the bra and ket sides, and its structure gives rise to path selectivity and interference between different pathways.
Substituting the Bogoliubov transformation, this bracket turns out to be
\begin{equation}
\begin{aligned}
&
\nu^{*}(\omega'_{\rm pu})
\mu(\omega'_{\rm St})
\nu^{*}(\omega'_{\rm pr})
\mu^{*}(\omega'_{\rm ga})
\\
&\qquad\times
\mu(\omega_{\rm ga})
\nu(\omega_{\rm pr})
\mu^{*}(\omega_{\rm St})
\nu(\omega_{\rm pu})
\sum_{j=1}^{18}\mathcal{D}_j^{\mathrm{ag}}
\\[1mm]
&+
\nu^{*}(\omega'_{\rm pu})
\mu(\omega'_{\rm St})
\nu^{*}(\omega'_{\rm pr})
\nu^{*}(\omega'_{\rm ga})
\\
&\qquad\times
\nu(\omega_{\rm ga})
\nu(\omega_{\rm pr})
\mu^{*}(\omega_{\rm St})
\nu(\omega_{\rm pu})
\sum_{j=1}^{6}\mathcal{D}_j^{\mathrm{gg}}.
\end{aligned}
\end{equation}
Here, each $\mathcal{D}_j$ is given by a product of four delta functions imposing the corresponding frequency constraints.
The 24 constituent terms are summarized in Table~\ref{tab:delta_terms}, where each row specifies one such product through the associated frequency constraints.
The terms $\mathcal{D}_j^{\rm gg}$ contain the factor
$\delta(\omega'_{\rm ga}-\omega_{\rm ga})$,
whereas $\mathcal{D}_j^{\rm ag}$ denotes the remaining terms.
The terms $\mathcal{D}_j^{\rm ag}$ represent contributions in which the gating mode is correlated with one of the modes constituting the anti-Stokes field, namely the pump, Stokes, or probe mode.
In contrast, for $\mathcal{D}_j^{\rm gg}$, the frequency of the gating mode is not correlated with the anti-Stokes field.

\begin{table*}[t]
\caption{
Delta-function products appearing in
$\mathcal{D}_{j}^{\rm ag}$ and $\mathcal{D}_{j}^{\rm gg}$.
The terms represented by $\mathcal{D}_{j}^{\rm ag}$ are classified
into the $D_{\rm diag}$, $D_{\rm off}$, $D_{\rm pg}$, and $D_{\rm pp}$ groups
according to their correlation structure.}
\label{tab:delta_terms}
\renewcommand{\arraystretch}{1}

\begin{ruledtabular}
\begin{tabular}{
@{}l@{\hspace{1.0em}}l@{\hspace{1.0em}}c@{\hspace{1.3em}}
l@{}
}
Class & Group & $j$ & Delta-function product \\
\hline

$\mathcal{D}_{j}^{\rm ag}$
& $D_{\rm diag}$ & 1 &
$\delta(\Omega_{e}-\omega'_{\rm pr}-\omega'_{\rm ga}) \delta(\Omega_{e}-\omega_{\rm pr}-\omega_{\rm ga})  \delta(\Omega_{e}-\omega'_{\rm St}-\omega_{\rm pu}) \delta(\Omega_{e}-\omega'_{\rm pu}-\omega_{\rm St})$
\\
&
& 2 &
$\delta(\Omega_{e}-\omega'_{\rm pr}-\omega'_{\rm ga})  \delta(\Omega_{e}-\omega_{\rm pr}-\omega_{\rm ga})  \delta(\omega'_{\rm St}-\omega_{\rm St})  \delta(\omega'_{\rm pu}-\omega_{\rm pu})$
\\
&
$D_{\rm off}$ & 3 &
$\delta(\Omega_{e}-\omega'_{\rm pr}-\omega'_{\rm ga})  \delta(\Omega_{e}-\omega_{\rm pu}-\omega_{\rm ga})  \delta(\Omega_{e}-\omega'_{\rm St}-\omega_{\rm pr})  \delta(\Omega_{e}-\omega'_{\rm pu}-\omega_{\rm St})$
\\
&
& 4 &
$\delta(\Omega_{e}-\omega_{\rm pr}-\omega_{\rm ga}) \delta(\Omega_{e}-\omega'_{\rm pu}-\omega'_{\rm ga}) \delta(\Omega_{e}-\omega'_{\rm pr}-\omega_{\rm St})  \delta(\Omega_{e}-\omega'_{\rm St}-\omega_{\rm pu})$
\\
&
& 5 &
$\delta(\Omega_{e}-\omega'_{\rm pr}-\omega'_{\rm ga})  \delta(\Omega_{e}-\omega_{\rm pu}-\omega_{\rm ga}) \delta(\omega'_{\rm St}-\omega_{\rm St})  \delta(\omega'_{\rm pu}-\omega_{\rm pr})$ 
\\
&
& 6 &
$\delta(\Omega_{e}-\omega_{\rm pr}-\omega_{\rm ga}) \delta(\Omega_{e}-\omega'_{\rm pu}-\omega'_{\rm ga})  \delta(\omega'_{\rm St}-\omega_{\rm St})  \delta(\omega'_{\rm pr}-\omega_{\rm pu})$
\\
&
& 7 &
$\delta(\Omega_{e}-\omega'_{\rm pr}-\omega'_{\rm ga})  \delta(\Omega_{e}-\omega'_{\rm St}-\omega_{\rm pr})  \delta(\omega_{\rm St}-\omega_{\rm ga})  \delta(\omega'_{\rm pu}-\omega_{\rm pu})$
\\
&
& 8 &
$\delta(\Omega_{e}-\omega_{\rm pr}-\omega_{\rm ga})
 \delta(\Omega_{e}-\omega'_{\rm pr}-\omega_{\rm St})
 \delta(\omega'_{\rm St}-\omega'_{\rm ga})
 \delta(\omega'_{\rm pu}-\omega_{\rm pu})$
\\
&
& 9 &
$\delta(\Omega_{e}-\omega'_{\rm pr}-\omega'_{\rm ga})
 \delta(\Omega_{e}-\omega'_{\rm St}-\omega_{\rm pu})
 \delta(\omega_{\rm St}-\omega_{\rm ga})
 \delta(\omega'_{\rm pu}-\omega_{\rm pr})$
\\
&
& 10 &
$\delta(\Omega_{e}-\omega_{\rm pr}-\omega_{\rm ga})
 \delta(\Omega_{e}-\omega'_{\rm pu}-\omega_{\rm St})
 \delta(\omega'_{\rm St}-\omega'_{\rm ga})
 \delta(\omega'_{\rm pr}-\omega_{\rm pu})$
\\

&
$D_{\rm pg}$ & 11 &
$\delta(\Omega_{e}-\omega'_{\rm pu}-\omega'_{\rm ga})
 \delta(\Omega_{e}-\omega_{\rm pu}-\omega_{\rm ga})
 \delta(\Omega_{e}-\omega'_{\rm St}-\omega_{\rm pr})
 \delta(\Omega_{e}-\omega'_{\rm pr}-\omega_{\rm St})$
\\
&
& 12 &
$\delta(\Omega_{e}-\omega'_{\rm pu}-\omega'_{\rm ga})
 \delta(\Omega_{e}-\omega'_{\rm St}-\omega_{\rm pr})
 \delta(\omega'_{\rm pr}-\omega_{\rm pu})
 \delta(\omega_{\rm St}-\omega_{\rm ga})$
\\
&
& 13 &
$\delta(\Omega_{e}-\omega_{\rm pu}-\omega_{\rm ga})
 \delta(\Omega_{e}-\omega'_{\rm pr}-\omega_{\rm St})
 \delta(\omega'_{\rm pu}-\omega_{\rm pr})
 \delta(\omega'_{\rm St}-\omega'_{\rm ga})$
\\
&
& 14 &
$\delta(\omega'_{\rm pu}-\omega_{\rm pr})
 \delta(\omega'_{\rm St}-\omega'_{\rm ga})
 \delta(\omega'_{\rm pr}-\omega_{\rm pu})
 \delta(\omega_{\rm St}-\omega_{\rm ga})$
\\
&
$D_{\rm pp}$ & 15 &
$\delta(\Omega_{e}-\omega'_{\rm pu}-\omega'_{\rm ga})
 \delta(\Omega_{e}-\omega'_{\rm St}-\omega_{\rm pu})
 \delta(\omega'_{\rm pr}-\omega_{\rm pr})
 \delta(\omega_{\rm St}-\omega_{\rm ga})$
\\
&
& 16 &
$\delta(\Omega_{e}-\omega_{\rm pu}-\omega_{\rm ga})
 \delta(\Omega_{e}-\omega'_{\rm pu}-\omega_{\rm St})
 \delta(\omega'_{\rm pr}-\omega_{\rm pr})
 \delta(\omega'_{\rm St}-\omega'_{\rm ga})$
\\
&
& 17 &
$\delta(\Omega_{e}-\omega'_{\rm pu}-\omega'_{\rm ga})
 \delta(\Omega_{e}-\omega_{\rm pu}-\omega_{\rm ga})
 \delta(\omega'_{\rm pr}-\omega_{\rm pr})
 \delta(\omega'_{\rm St}-\omega_{\rm St})$
\\
&
& 18 &
$\delta(\omega'_{\rm pu}-\omega_{\rm pu})
 \delta(\omega'_{\rm St}-\omega'_{\rm ga})
 \delta(\omega'_{\rm pr}-\omega_{\rm pr})
 \delta(\omega_{\rm St}-\omega_{\rm ga})$
\\

\hline

$\mathcal{D}_{j}^{\rm gg}$
& & 1 &
$\delta(\omega'_{\rm ga}-\omega_{\rm ga})
 \delta(\Omega_{e}-\omega'_{\rm St}-\omega_{\rm pu})
 \delta(\Omega_{e}-\omega'_{\rm pu}-\omega_{\rm St})
 \delta(\omega'_{\rm pr}-\omega_{\rm pr})$
\\
&
& 2 &
$\delta(\omega'_{\rm ga}-\omega_{\rm ga})
 \delta(\Omega_{e}-\omega'_{\rm St}-\omega_{\rm pr})
 \delta(\Omega_{e}-\omega'_{\rm pr}-\omega_{\rm St})
 \delta(\omega'_{\rm pu}-\omega_{\rm pu})$
\\
&
& 3 &
$\delta(\omega'_{\rm ga}-\omega_{\rm ga})
 \delta(\Omega_{e}-\omega'_{\rm St}-\omega_{\rm pr})
 \delta(\Omega_{e}-\omega'_{\rm pu}-\omega_{\rm St})
 \delta(\omega'_{\rm pr}-\omega_{\rm pu})$
\\
&
& 4 &
$\delta(\omega'_{\rm ga}-\omega_{\rm ga})
 \delta(\Omega_{e}-\omega'_{\rm St}-\omega_{\rm pu})
 \delta(\Omega_{e}-\omega'_{\rm pr}-\omega_{\rm St})
 \delta(\omega'_{\rm pu}-\omega_{\rm pr})$
\\
&
& 5 &
$\delta(\omega'_{\rm ga}-\omega_{\rm ga})
 \delta(\omega'_{\rm St}-\omega_{\rm St})
 \delta(\omega'_{\rm pu}-\omega_{\rm pr})
 \delta(\omega'_{\rm pr}-\omega_{\rm pu})$
\\
&
& 6 &
$\delta(\omega'_{\rm ga}-\omega_{\rm ga})
 \delta(\omega'_{\rm St}-\omega_{\rm St})
 \delta(\omega'_{\rm pu}-\omega_{\rm pu})
 \delta(\omega'_{\rm pr}-\omega_{\rm pr})$
\end{tabular}
\end{ruledtabular}
\end{table*}

We classify the contributions $\mathcal{D}_j^{\rm ag}$ according to whether they contain both $\delta(\Omega_e-\omega_{\rm pr}-\omega_{\rm ga})$ and its primed counterpart, only one of them, or neither. 
The first group ($D_{\rm diag}$) involves identical bra- and ket-side pathways, whereas the second ($D_{\rm off}$) involves interference between distinct pathways. 
The terms containing neither are further divided into $D_{\rm pg}$, which involves correlations between the CARS pump and SFG gating fields, and $D_{\rm pp}$, which involves correlations between the bra- and ket-side probe fields. 
We define $D_{\rm diag}=\sum_{j=1}^{2}\mathcal{D}_j^{\rm ag}$, $D_{\rm off}=\sum_{j=3}^{10}\mathcal{D}_j^{\rm ag}$, $D_{\rm pg}=\sum_{j=11}^{14}\mathcal{D}_j^{\rm ag}$, and $D_{\rm pp}=\sum_{j=15}^{18}\mathcal{D}_j^{\rm ag}$. 
Accordingly, the total SFG spectrum is decomposed into the sum of these contributions, $S = S_{\rm diag} + S_{\rm off} + S_{\rm pg} + S_{\rm pp} + S_{\rm gg},$
where each subscript on $S$ denotes the spectral contribution obtained by integrating Eq.~(\ref{eq:SFG_spectrum}) with the corresponding delta-function group.

The contributions from $D_{\rm pp}$ and $\mathcal{D}_j^{\rm gg}$ can be subtracted from the total SFG output. 
$D_{\rm pp}$ represents a response independent of $\tau_{\rm pr}$
because the delta function $\delta(\omega'_{\rm pr}-\omega_{\rm pr})$ sets the phase factor $e^{-i(\omega'_{\rm pr}-\omega_{\rm pr})\tau_{\rm pr}}$ to unity in Eq.~\ref{eq:SFG_spectrum}. 
Evaluating the SFG output at a delay $\tau_{\rm pr}$ chosen to be much longer than both the temporal correlation width of the twin beams and the material relaxation time allows $D_{\rm pp}$ to be isolated. 
Similarly, $\mathcal{D}_j^{\rm gg}$ is isolated by setting $\tau_{\rm ga}$ to a sufficiently long delay. 
In the following, we focus on the terms other than $D_{\rm pp}$ and $\mathcal{D}_j^{\rm gg}$.

We derive the analytical expression for $S_{\rm diag}$.
For simplicity, we apply the high-gain approximation $\mu(\omega) \approx \nu(\omega)$ \cite{PhysRevA.87.013844}. 
We introduce $M(\omega) \equiv \mu^2(\omega)$, which physically represents the spectral intensity distribution of the twin beams.
After an appropriate change of variables, it can be written as
\begin{equation}
\begin{aligned}
S_{{\rm diag}}&(\omega_{\rm \Delta}, \tau_{\rm pr}, \tau_{\rm ga}) = 
2 \left| A(\tau_{\rm pr}, \tau_{\rm ga}) \right|^2 
\left| \chi(\omega_{\rm \Delta}) \right|^2 \\
&\times \int_{\frac{\Omega_{e}}{2}}^{\omega_{\rm \Delta}+\frac{\Omega_{e}}{2}} d\omega_{l}
M(\omega_{l})M(\omega_{\rm \Delta}+\Omega_{e}-\omega_{l}),
\end{aligned}
\label{eq:Sdiag}
\end{equation}
where $\omega_{\rm \Delta}=\omega_{\rm sfg}-\Omega_{e}$ and 
\begin{equation}
\begin{aligned}
&A(\tau_{\rm pr},\tau_{\rm ga}) =\int_{\Omega_e/2}^{\Omega_e} d\omega_k\,M(\omega_k) e^{i\omega_k(\tau_{\rm ga}-\tau_{\rm pr})}.
\end{aligned}
\label{eq:A}
\end{equation}

In $S_{\rm diag}$, the probe and gating fields are correlated on both the ket and bra sides due to the $D_{\rm diag}$ constraint.
Consequently, the arguments of both $\chi$ and $\chi^*$ reduce to $\omega_\Delta = \omega_{\rm sfg} - \Omega_e$, establishing a direct one-to-one correspondence between the Raman susceptibility $|\chi(\omega_\Delta)|^2$ and the SFG output profile. 
Conversely, because $|\chi(\omega_\Delta)|^2$ factors out of the frequency integral, the temporal relaxation of the Lorentzian line shape cannot be observed by scanning the delays.

Importantly, the spectral convolution with the probe field, which degrades the spectral resolution in conventional CARS spectroscopy, is effectively removed in $S_{\rm diag}$. 
This occurs because the sum frequency of the correlated probe and gating fields is restored to the PDC excitation frequency, $\omega_{\rm pr}+\omega_{\rm ga}=\Omega_e$, through pathway enhancement under the $D_{\rm diag}$ constraint. 
The loss of temporal information regarding relaxation can also be understood as a consequence of the probe and gating fields being correlated in both the bra and ket sides. 
Because the twin-beam correlation is a two-mode correlation, the probe becomes unable to form temporal correlations with any other modes in such a condition.


Our main interest is whether contributions involving probe–gating correlations that eliminate probe-induced spectral convolution on either the bra or ket side can retain information about Raman relaxation dynamics. 
We therefore analyze $S_{\rm off}$, which can be expressed as
\begin{equation}
\begin{aligned}
&S_{\rm off}(\omega_\Delta,\tau_{\rm pr},\tau_{\rm ga})\\
&=4A(\tau_{\rm pr},\tau_{\rm ga})
\chi^*(\omega_\Delta)
X(\omega_\Delta,\tau_{\rm pr},\tau_{\rm ga})
+\mathrm{c.c.},
\end{aligned}
\end{equation}
where
\begin{equation}
\begin{aligned}
&X(\omega_\Delta,\tau_{\rm pr},\tau_{\rm ga}) = \int_{\frac{\Omega_e}{2}}^{\omega_\Delta+\frac{\Omega_e}{2}} d\omega_l \int_{\frac{\Omega_e}{2}}^{\Omega_e} d\omega_m 
\\ & 
\quad\times \chi(\omega_m-\omega_l+\omega_\Delta) M(\omega_m)M(\omega_l)M(\omega_\Delta+\Omega_e-\omega_l) 
\\ &\quad\times
e^{i\omega_l\tau_{\rm pr}} e^{-i\omega_m\tau_{\rm ga}}.
\end{aligned}
\end{equation}
As $\tau_{\rm pr}$ is varied, the factor $A$ decreases owing to the temporal correlation between the CARS probe and SFG gating fields. 
To eliminate this effect, we set $\tau_{\rm pr}=\tau_{\rm ga}=\tau$. 
Since $A$ becomes a real constant, we only need to examine whether $\chi^*(\omega_\Delta)X+\mathrm{c.c.}$ contains information about Raman relaxation dynamics.

For further simplification, we assume a Gaussian spectral profile for the signal beam of the twin beams, 
$M(\omega)=\exp\!\left[ -\frac{(\omega-\Omega_{\rm sig})^2}{2\sigma^2} \right]$,
with center frequency $\Omega_{\rm sig}=(\Omega_e+\Omega_R)/2$ and spectral width $\sigma$. 
Thus, the difference between the center frequencies of the signal and idler twin beams matches $\Omega_R$. 
We set the SFG frequency to $\omega_{\rm sfg}=\Omega_e+\omega_\Delta=\Omega_e+\Omega_R$.

We express the Raman susceptibility in the time domain as 
$\chi(\omega)=\frac{1}{(\Omega_R-\omega)-i\Gamma} = i\int_0^\infty dt e^{-\Gamma t}e^{-i(\Omega_R-\omega)t}$.
Substituting this representation into $X$ and changing the integration variables to $x=\omega_l-\Omega_{\rm sig}$ and $y=\omega_m-\Omega_{\rm sig}$, we obtain
\begin{equation}
\begin{aligned}
&\chi^*(\Omega_R)X(\Omega_R,\tau,\tau)=\frac{1}{\Gamma}\int_0^\infty dt\, e^{-\Gamma t} 
F(\tau-t)
G(\tau-t),
\end{aligned}
\end{equation}
where
\begin{equation}
\begin{aligned}
F(\tau-t) =\int_{-\Omega_R/2}^{\Omega_R/2}dx\,  e^{-\frac{x^2}{\sigma^2}}e^{ix(\tau-t)}
\end{aligned}
\end{equation}
and
\begin{equation}
\begin{aligned}
G(\tau-t) =\int_{-\Omega_R/2}^{(\Omega_e-\Omega_R)/2}dy\,  e^{-\frac{y^2}{2\sigma^2}}e^{-iy(\tau-t)}.
\end{aligned}
\end{equation}

When the integration ranges in $F(\tau-t)$ and $G(\tau-t)$ can be extended to $(-\infty,\infty)$, both functions reduce to Gaussian temporal profiles, analogous to the electric-field envelopes of ultrashort pulses.
This approximation is valid when $\sigma \ll \min(\Omega_R/2,(\Omega_e-\Omega_R)/2)$. 
Scanning $\tau$ then changes the temporal overlap between Raman excitation and probing, as described by the product $F(\tau-t)G(\tau-t)$. 
The integration over $t$, weighted by the exponential factor $e^{-\Gamma t}$, accounts for the Raman relaxation history sampled through this temporal overlap. 
Consequently, the real part of $\chi^*(\Omega_R)X(\Omega_R,\tau,\tau)$ contains information about Raman relaxation dynamics. 
When the temporal overlap is sufficiently short compared with the Raman relaxation time, this delay dependence approaches the exponential decay $e^{-\Gamma\tau}$.

The analysis of $S_{\rm off}$ shows that $\chi^*(\omega_\Delta)$ provides a Raman spectral response free from convolution with the probe spectrum, while $X$ retains time-resolved information about Raman relaxation. 
This contribution therefore exhibits signatures of both narrowband- and ultrashort-pulse-like probing. However, experimentally isolating this contribution remains challenging, particularly because it must be distinguished from $S_{\rm diag}$ and $S_{\rm ap}$. 
Addressing this issue is left for future work.

In summary, we have theoretically investigated CARS driven by bright twin beams with SFG readout. We have identified the physical origin of the high spectral and temporal resolution carried by specific contributions to the SFG output. 
Frequency correlations between the probe and gating fields yield spectral contributions that reflect the intrinsic Raman line shape without probe-induced convolution, while interference between distinct bra- and ket-side pathways enables time-resolved access to Raman relaxation dynamics. 
Our results thus clarify the correlation mechanisms underlying high-resolution Raman spectroscopy with bright twin beams and the practical limitations on their use.

\paragraph*{Aknowledgments}
We would like to express our gratitude to the HONMAMON co-creative research program between HORIBA and Kyoto University.
We thank Dr. Y. Fujihashi for bringing relevant previous studies to our attention.
This work was supported by the Japan Science and Technology Agency (JST) PRESTO (Grant No. JPMJPR2352), the Japan Society for the Promotion of Science (JSPS) KAKENHI (Grant Nos. 26K21749, 26K22832 and 24K06919), and the RIKEN Transformative Research Innovation Platform (TRIP) initiative. 
We acknowledge the use of AI assistance, specifically ChatGPT, in refining the language and clarity of this manuscript.



\bibliography{apssamp}

\end{document}